\documentclass[10pt,letterpaper,twoside]{article}

\usepackage{times}
\usepackage[dvips]{graphicx}
\usepackage{fancyhdr}
\usepackage{multicol}
\usepackage{ifthen}

\usepackage{amsmath}
\usepackage{amssymb}
\usepackage{natbib}
\bibpunct[, ]{(}{)}{;}{a}{}{,}
\usepackage{setspace} 
\usepackage{url}
\newcommand{\jamesversion}{discussion}   
\newcommand{\jamesfulltitle}
           {Atmospheric dynamics of Earth-like tidally locked aquaplanets}
\newcommand{\jamesshorttitle}   
           {Tidally locked Earth}
\newcommand{\jamesfullauthors}   
           {Timothy M. Merlis$\;^1$ and Tapio Schneider$\;^1$}
\newcommand{\jamesshortauthors}   
           {Merlis and Schneider}
\newcommand{\jamesaffiliations}   
           {$^1$California Institute of Technology}
\newcommand{\jamescorraddress}   
           {Timothy M. Merlis, 1200 E. California Blvd. MC 100-23, Pasadena, CA 91125 \\
            e-mail: tmerlis@caltech.edu}
\newcommand{\jamessubmitdate}{26 January 2010}   
\newcommand{\jamesrevisedate}{}                

\newlength{\lpgmargin} 
\newlength{\rpgmargin} 
\newlength{\tpgmargin} 
\newlength{\bpgmargin} 
                                                                                
\newlength{\netwidth}
\newlength{\colwidth}
\newlength{\rulespace}
\newcommand{\jamesshortsuffix}
           {\ifthenelse{\equal{\jamesversion}{final}}{}{-D}}
\newcommand{\jameslongsuffix}
           {\ifthenelse{\equal{\jamesversion}{final}}{}{ -- Discussion}}
\newcommand{\jameslogo}
           {\ifthenelse{\equal{\jamesversion}{final}}
              {\includegraphics[height=1.0in]{JAMES_logo.eps}}
              {\includegraphics[height=1.0in]{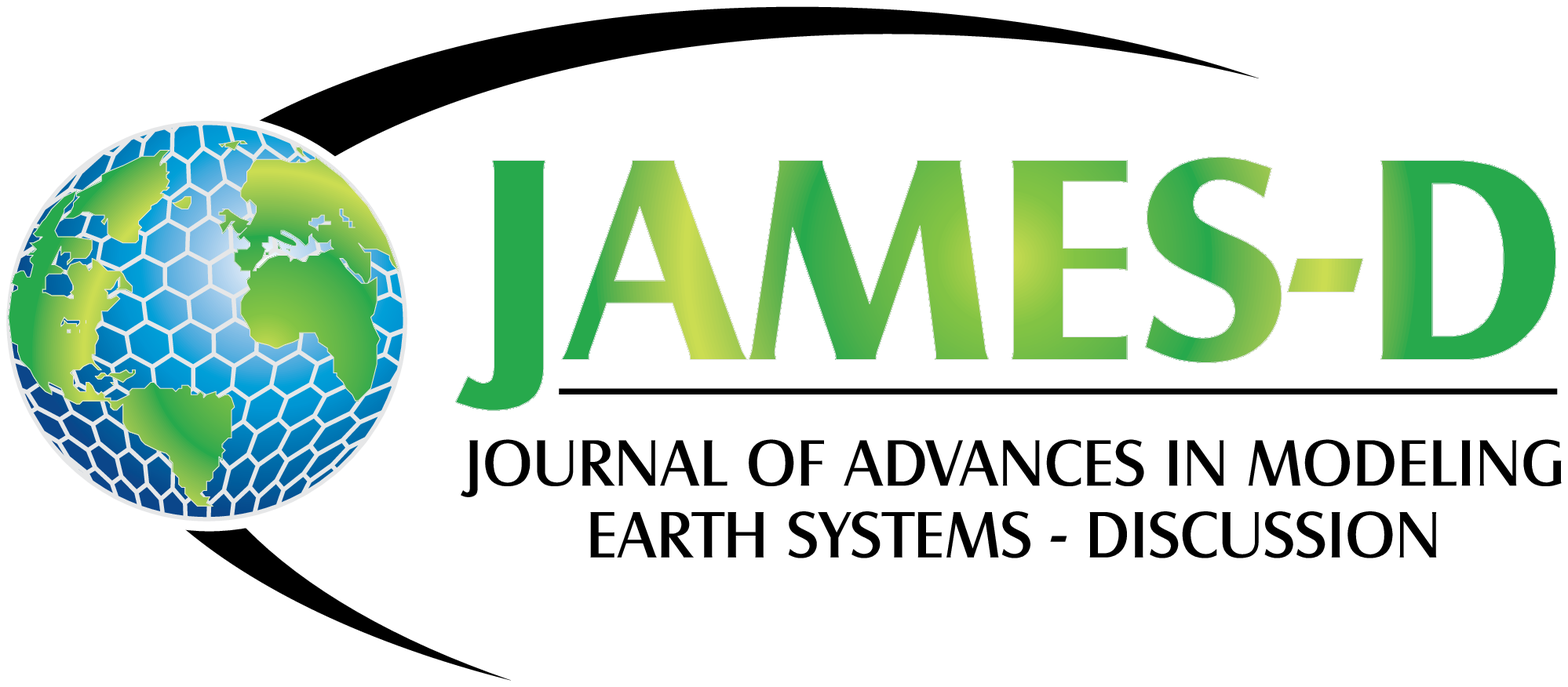}}}

\fancypagestyle{plain}{%
\fancyhf{}
\fancyhead[RO,LE]{\textsf{\thepage}}
\fancyfoot[CO]{\textsf{\emph{Journal of Advances in Modeling Earth Systems\jameslongsuffix}}}
\fancyfoot[CE]{\textsf{\emph{JAMES\jamesshortsuffix}}}

}

\makeatletter
\renewcommand{\@makefntext}[1]
   {\noindent{\@makefnmark}#1}
\makeatother

\makeatletter
\renewcommand{\@seccntformat}[1]{{\csname the#1\endcsname}.\hspace{0.75em}}
\renewcommand{\section}{\@startsection
   {section}{1}{0mm}{-1.5\baselineskip}{0.5\baselineskip}%
   {\sffamily\normalsize\bfseries}}
\renewcommand{\subsection}{\@startsection
   {subsection}{2}{0mm}{-1.5\baselineskip}{0.5\baselineskip}%
   {\sffamily\normalsize\itshape}}

\makeatother

\numberwithin{equation}{section}

\def\hang1{\smallskip\noindent\hangindent=.25truein\hangafter=1}

\newcommand  {\about} {\mathop{\sim}\!}
\newcommand  {\Ro} {\mathrm{Ro}}
\newcommand  {\Fr} {\mathrm{Fr}}
\urldef{\slow_anim}\url{http://www.gps.caltech.edu/~tapio/animations/tidally_locked_slow.mov}
\urldef{\fast_anim}\url{http://www.gps.caltech.edu/~tapio/animations/tidally_locked_fast.mov}

\begin{document}

\thispagestyle{plain}
\begin{flushleft}
\vspace*{-1.0in}
\jameslogo
\\[0.2in]

{\Large\textsf{\textbf{\jamesfulltitle}}}
\\[0.2in]

{\large\textsf{\emph{\jamesfullauthors}}}
\\[0.2in]

{\small\textsf{\jamesaffiliations}}

\renewcommand{\thefootnote}{}
\footnotetext{\textsf{\textbf{To whom correspondence should be addressed.} \\
\jamescorraddress}}
\renewcommand{\thefootnote}{\arabic{footnote}}

\rule{\textwidth}{0.5pt}
\\[0.1in]

{\small\textsf{Manuscript submitted \jamessubmitdate
  {\ifthenelse{\equal{\jamesrevisedate}{}}{}
     {, revised \jamesrevisedate}}
}}
\end{flushleft}

\noindent{\small %
  We present simulations of atmospheres of Earth-like aquaplanets that are
  tidally locked to their star, that is, planets whose orbital period
  is equal to the rotation period about their spin axis, so that one
  side always faces the star and the other side is always dark. As
  extreme cases illustrating the effects of slow and rapid rotation,
  we consider planets with rotation periods equal to one current Earth
  year and one current Earth day. The dynamics responsible for the
  surface climate (e.g., winds, temperature, precipitation) and the
  general circulation of the atmosphere are discussed in light of
  existing theories of atmospheric circulations. For example, as
  expected from the increasing importance of Coriolis accelerations
  relative to inertial accelerations as the rotation rate increases,
  the winds are approximately isotropic and divergent at leading order
  in the slowly rotating atmosphere but are predominantly zonal and
  rotational in the rapidly rotating atmosphere. Free-atmospheric
  horizontal temperature variations in the slowly rotating atmosphere
  are generally weaker than in the rapidly rotating
  atmosphere. Interestingly, the surface temperature on the night
  side of the planets does not fall below $\about 240\,\mathrm{K}$
  in either the rapidly or slowly rotating atmosphere;
  that is, heat transport from the day side to the night side of the
  planets efficiently reduces temperature contrasts in either
  case. Rotational waves and eddies shape the distribution of winds,
  temperature, and precipitation in the rapidly rotating atmosphere;
  in the slowly rotating atmosphere, these distributions are
  controlled by simpler divergent circulations.  The results are of
  interest in the study of tidally locked terrestrial exoplanets and
  as illustrations of how planetary rotation and the insolation
  distribution shape climate.} \\

\begin{multicols}{2}   


\section{Introduction}

Planets generally evolve toward a state in which they become tidally
locked to their star. Torques the star exerts on tidal bulges on a
planet lead to an exchange between the spin angular momentum of the
planetary rotation and orbital angular momentum of the planet's
revolution around the star, such that the rotation period around the
spin axis gradually approaches the orbital period of the planet
\citep{hubbard84}. (The spin angular momentum of the star may also
participate in this angular momentum exchange.) This process reaches
its tidally locked end state when the rotation period is equal to the
orbital period, so that one side of the planet always faces the star
and the other side is always dark. The time it takes to reach this end
state may exceed the lifetime of the planetary system, so it may never
be reached (this is almost certainly the case for the Sun-Earth
system). But planets that are close to their star can reach a tidally
locked state more quickly. Such close planets in other solar systems
are easier to detect than planets farther away from their star, and
exoplanets that are believed to be tidally locked have indeed been
detected in recent years \citep[e.g.,][]{Charbonneau00}.  Here we
investigate the atmospheric dynamics of Earth-like tidally locked
aquaplanets through simulations with a three-dimensional general
circulation model (GCM). Our purpose is pedagogic: we contrast rapidly
and slowly rotating tidally locked Earth-like planets with each other
and with Earth itself to illustrate the extent to which atmospheric
dynamics depend on the insolation distribution and planetary rotation
rate.

There are two areas of existing research on the atmospheric dynamics
of tidally locked planets. First, there are several studies motivated
by ``hot Jupiters''---large, close-in planets that have been observed
transiting a star (see \citet{showman09} for a review). Second, there
are studies of Earth-like exoplanets closely orbiting a relatively
cool star---planets that have not yet been but may be detected soon,
for example, by NASA's recently launched \emph{Kepler}
space telescope or by ground-based telescopes 
\citep[as demonstrated by the observations of][]{charbonneau09}. 
\citet{joshi97} investigated the large-scale
circulation of such Earth-like planets and explored how their climate
depends on the mass of the atmosphere. And \citet{joshi03} documented
their hydrological cycle using an Earth climate model.

The studies by Joshi et al.\ provide a description of Earth-like
tidally locked atmospheric circulations and how they depend on some
parameters, such as the atmospheric mass. But some questions remain,
among them:
(i) How does the planet's rotation rate affect the circulation and
climate?  (ii) What controls the precipitation distribution (location
of convergence zones)?  (iii) What mechanisms generate large-scale
features of the circulation, such as jets?  (iv) What determines the
atmospheric stratification?

We address these questions by simulating tidally locked Earth-like
aquaplanets with rotation periods equal to one current Earth year and one
current Earth day. These are merely two illustrative cases: The slowly
rotating case corresponds roughly to the tidally locked end state
Earth would reach if the sun were not changing, that is, if the solar
constant $S_0 = 1367 \ \mathrm{W m}^{-2}$ remained fixed. The rapidly
rotating case corresponds to a terrestrial planet sufficiently close
to a cool host star, so that the orbital period is one Earth day but
the average insolation reaching the planet still is $S_0 = 1367 \
\mathrm{W m}^{-2}$, as presently on Earth.

\section{General circulation model}

We use an idealized atmosphere GCM with an active hydrological cycle
and an aquaplanet (slab ocean) lower boundary condition. 
The slab ocean has no explicit horizontal transports but is implicitly
assumed to transport water from regions of net precipitation to 
regions of net
evaporation, so that the local water supply is unlimited. The GCM
is a modified version of the model described in \citet{ogorman08b},
which is similar to the model described in
\citet{frierson06a}. Briefly, the GCM is a three-dimensional
primitive-equation model of an ideal-gas atmosphere. It uses the
spectral transform method in the horizontal, with resolution T85, and
finite differences in $\sigma=p/p_s$ coordinates (pressure $p$ and
surface pressure $p_s$) in the vertical, with $30$ unequally spaced
levels. Subgrid-scale dissipation is represented by an exponential
cutoff filter \citep{smith02b}, which acts on spherical wavenumbers
greater than 40, with a damping timescale of $1\,\mathrm{h}$ on the
smallest resolved scale. Most features of the simulated flows are
similar at T42 resolution or with different subgrid-scale filters;
exceptions are noted below.

The GCM has a surface with uniform albedo ($0.38$) and uniform
roughness lengths for momentum fluxes ($5 \times 10^{-3}\,\mathrm{m}$)
and for latent and sensible heat fluxes ($10^{-5}\,\mathrm{m}$). It
has a gray radiation scheme, and a quasi-equilibrium moist convection
scheme that relaxes convectively unstable atmospheric columns to a
moist pseudo-adiabat with constant relative humidity $70\%$
\citep{frierson07b}. Only the vapor-liquid phase transition is taken
into account, and the latent heat of vaporization is taken to be
constant. No liquid water is retained in the atmosphere, so any
precipitation that forms immediately falls to the surface. Radiative
effects of clouds are not taken into account, except insofar as the
surface albedo and radiative parameters are chosen to mimic some of
the global-mean effects of clouds on Earth's radiative budgets.

Other model details can be found in \citet{ogorman08b}. However, the
radiation scheme differs from that in \citet{ogorman08b} and is
described in what follows.

\subsection{Tidally locked insolation}

The top-of-atmosphere insolation is held fixed at the instantaneous
value for a spherical planet \citep[e.g.,][]{hartmann94b},
\begin{equation}
  S_{\mathrm{TOA}} = S_0 \times \mathrm{max}\bigl(0, \cos(\phi) \cos(\lambda - \lambda_0)\bigr),
\end{equation}
where $\phi$ is latitude and $\lambda$ is longitude, with subsolar
longitude $\lambda_0 = 270^{\circ}$ and solar constant $S_0 = 1367 \
\mathrm{W m}^{-2}$.  There is no diurnal cycle of insolation.
That is, we assume zero eccentricity of the orbit and zero obliquity
of the spin axis.

\subsection{Longwave optical depth}

\citet{ogorman08b} used a gray radiation scheme with a longwave
optical depth that was uniform in longitude and a prescribed function
of latitude, thereby ignoring any longwave water vapor feedback. This
is not adequate for tidally locked simulations with strong radiative
variations in longitude. Therefore, we use a longwave optical depth
that varies with the local water vapor concentration
\citep[cf.][]{thuburn00}, providing a crude representation of longwave
water vapor feedback.

As in \citet{frierson06a}, the longwave optical depth $\tau$ has a
term linear in pressure $p$, representing well-mixed absorbers like
CO$_2$, and a term quartic in pressure, representing water vapor, an
absorber with a scale height that is one quarter of the pressure scale
height,
\begin{equation}
  \tau = \tau_0 \left( \frac{p}{p_0} \right)^4 + \tau_1 \left( \frac{p}{p_0} \right).
\end{equation}
Here, $\tau_1 = 1.2$ and $p_0 = 10^5\,\mathrm{Pa}$ are
constants. However, the optical thickness of water vapor is a function
of the model's instantaneous column water vapor concentration,
\begin{equation}
  \tau_0 = \frac{1}{p_1} \int_{0}^{p_s} q \, dp,
\end{equation}
with specific humidity $q$ and an empirical constant $p_1 = 980
\,\mathrm{Pa}$ to keep $\tau_0$ order one for conditions typical of
Earth's tropics.

The details of the radiation scheme such as the constants chosen
affect quantitative aspects of the simulations (e.g., the precise
surface temperatures obtained), but not the large-scale dynamics on
which we focus.  The simulated surface climate is similar to those
presented in \citet{joshi03}, who used a more complete
representation of radiation including radiative effects of clouds.
This gives us confidence that the qualitative results presented here do not 
depend on the details of the radiation scheme.

\subsection{Simulations}

We conducted a rapidly rotating simulation with planetary rotation
rate equal to that of present-day Earth, $\Omega = \Omega_E = 7.292
\times 10^{-5} \mathrm{s}^{-1}$, and a slowly rotating simulation with
planetary rotation rate approximately equal to one present-day Earth
year, $\Omega=\Omega_E/365$.

The results we present are averages over the last 500~days of 4000-day
simulations (with 1~day = $86400\,\mathrm{s}$, irrespective of the
planetary rotation rate). During the first 3000~days of the
simulation, we adjusted the subgrid-scale dissipation parameters to
the values stated above to ensure there was no energy build-up at the
smallest resolved scales.

\section{Slowly rotating simulation}

The Rossby number $\Ro = U/(fL)$, with horizontal velocity scale $U$,
Coriolis parameter $f$, and length scale $L$, is a measure of the
importance of inertial accelerations ($\about U^2/L$) relative to
Coriolis accelerations ($\about f U$) in the horizontal momentum
equations. In rapidly rotating atmospheres, including Earth's, the
Rossby number in the extratropics is small, and the dominant balance
is geostrophic, that is, between pressure gradient forces and Coriolis
forces. In slowly rotating atmospheres, the Rossby number may not be
small. If $\Ro= O(1)$, inertial and Coriolis accelerations both are
important, as in the deep tropics of Earth's atmosphere. If $\Ro \gg
1$, Coriolis accelerations and effects of planetary rotation become
unimportant. In that case, there is no distinguished direction in the
horizontal momentum equations, so the horizontal flow is expected to
be isotropic, that is, the zonal velocity scale $U$ and meridional
velocity scale $V$ are of the same order. This is the dynamical regime
of our slowly rotating simulation.

In this dynamical regime, the magnitude of horizontal temperature
variations can be estimated through scale analysis of the horizontal
momentum equations \citep{charney63}. In the free atmosphere, where
frictional forces can be neglected, inertial accelerations scale
advectively, like $\about U^2/L$, and are balanced by accelerations
owing to pressure gradients, which scale like $\delta p / (\rho L)$,
where $\delta p$ is the scale of horizontal pressure variations and
$\rho$ is density.  The density and vertical pressure variations are
related by the hydrostatic relation, $p / H \sim \rho g$, where
$H=RT/g$ is the pressure scale height (specific gas constant $R$ and
temperature $T$). If one combines the scalings from the horizontal
momentum and hydrostatic equations, horizontal variations in pressure,
density, and (potential) temperature (using the ideal gas law) scale
like
\begin{equation}
\frac{\delta p }{p} \sim \frac{\delta \rho }{\rho} \sim 
\frac{\delta \theta }{\theta} \sim \frac{U^2}{g H} \equiv \Fr,
\end{equation}
where $\Fr = U^2/(gH)$ is the Froude number. For a terrestrial planet
with $H \approx 7\,\mathrm{km}$ and $g=9.8\,\mathrm{m\,s^{-2}}$, the
Froude number is $\Fr \lesssim 10^{-2}$ for velocities $U \lesssim
25\,\mathrm{m\,s^{-1}}$. So free-atmospheric horizontal temperature
and pressure variations are expected to be small insofar as velocities
are not too strong (e.g., if their magnitude is limited by shear 
instabilities). This is the case in the tropics of Earth's atmosphere,
and these expectations are also borne out in the slowly rotating simulation.

\subsection{Surface temperature}

\begin{figure*}[!tbh]
\centerline{ \includegraphics[width=6.0in,clip=true]{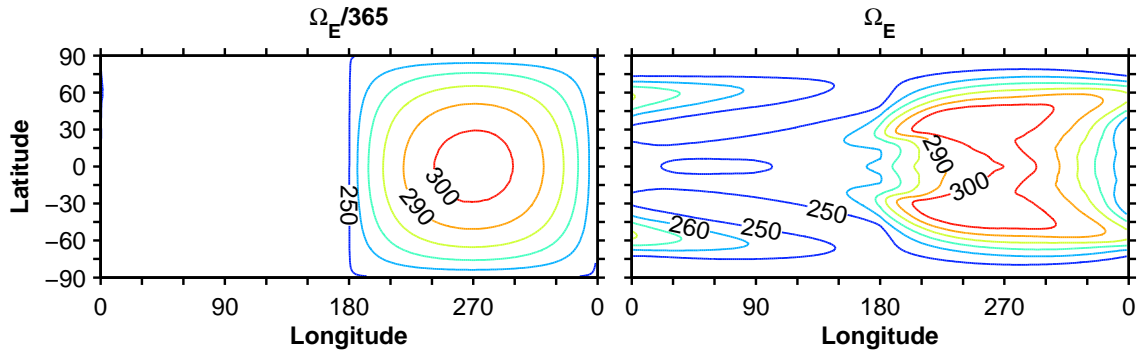} }
\caption{Surface temperature (K) in simulations with $\Omega_E/365$ (left) and
  $\Omega_E$ (right). The contour interval is
  $10\,\mathrm{K}$. }
\label{fig-surf_temp}
\end{figure*}
In the slowly rotating simulation, the surface temperature mimics the
insolation distribution on the day side of the planet, decreasing
monotonically and isotropically away from the subsolar point; the
night side of the planet has a nearly uniform surface temperature
(Fig.~\ref{fig-surf_temp}). Though the surface temperature resembles
the insolation distribution, the influence of atmospheric dynamics is
clearly evident in that the night side of the planet is considerably
warmer than the radiative-convective equilibrium temperature of
$0\,\mathrm{K}$. 

\subsection{Hydrological cycle}

\begin{figure*}[!tbh]
\centerline{ \includegraphics[width=6.0in,clip=true]{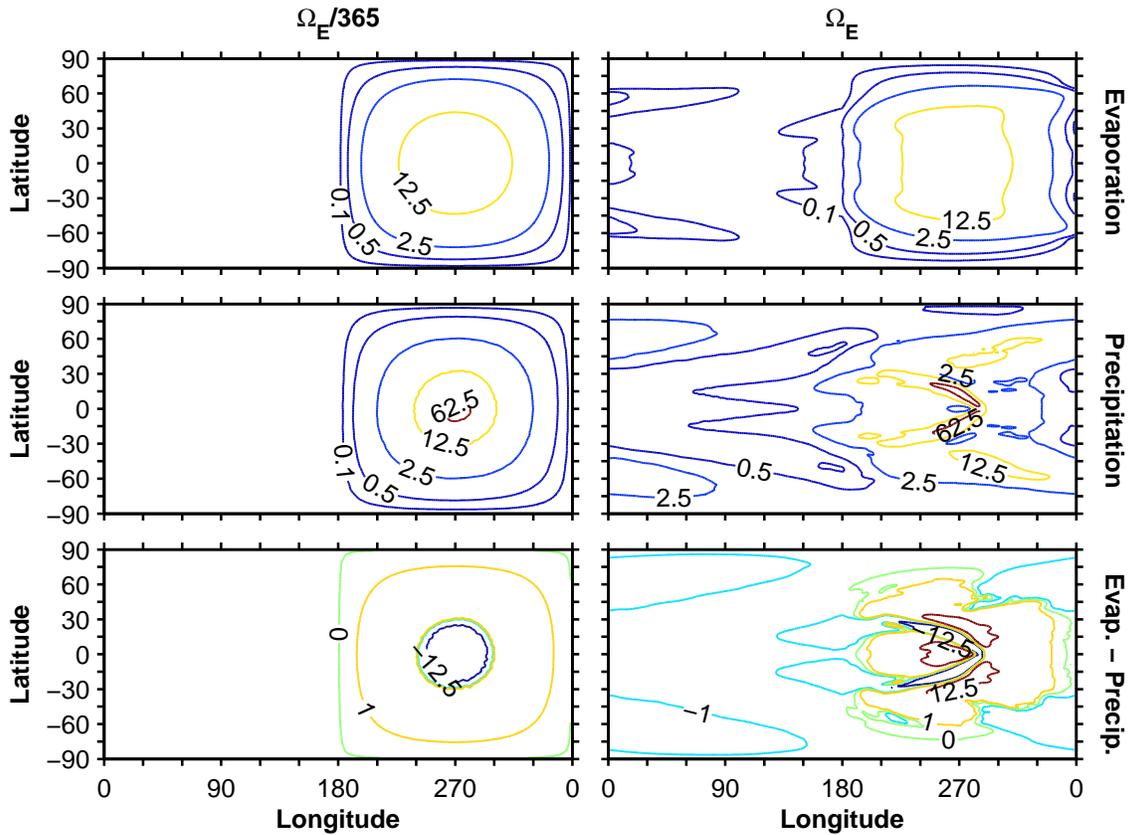} }
\caption{Evaporation (top row, mm/day), precipitation (middle row,
  mm/day), and evaporation minus precipitation (bottom row, mm/day) in
  simulations with $\Omega_E/365$ (left) and $\Omega_E$ (right). 
  Contours are shown at $0.1 \times 5^{0, 1, 2, 3, 4} \, \mathrm{mm\,day^{-1}}$
  in the top two rows and $-12.5, -1.0, 0, 1.0, 12.5 \, \mathrm{mm\,day^{-1}}$
  in the bottom row.}
\label{fig-hydro}
\end{figure*}
\begin{figure*}[!tbh]
\centerline{ \includegraphics[width=3.0in,clip=true]{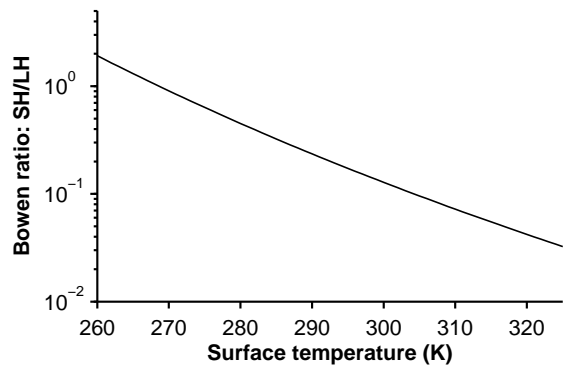} }
\caption{Bowen ratio vs surface temperature assuming a 3-K
  surface--air temperature difference, 70\% relative humidity, and
  surface and surface air pressure of $10^5$ Pa,
  using the same simplified saturation vapor pressure formula as in
  the GCM. 
  }
\label{fig-bowen}
\end{figure*}

Surface evaporation rates likewise mimic the insolation distribution
(Fig.~\ref{fig-hydro}), as would be expected from a surface energy
budget in which the dominant balance is between heating by shortwave
radiation and cooling by evaporation. This is the dominant balance
over oceans on Earth \citep[e.g.,][]{Trenberth09a}, and in general on
sufficiently warm Earth-like aquaplanets
\citep{Pierrehumbert02,ogorman08b}. The dominance of evaporation in
the surface energy budget in sufficiently warm climates can be
understood by considering how the Bowen ratio, the ratio of sensible
to latent surface fluxes, depends on temperature. For surface fluxes
given by bulk aerodynamic formulas, the Bowen ratio, $B$,
depends on the surface temperature, $T_s$, the near-surface air
temperature, $T_a$, and the near-surface relative humidity,
$\mathcal{H}$,
\begin{equation}
  B \sim \frac{c_p (T_s - T_a)}{L \bigl( q^*(T_s) - \mathcal{H} q^*(T_a) \bigr)},
\end{equation}
where $q^*$ is the saturation specific humidity, $c_p$ is the
specific heat of air at constant pressure, and $L$ is the latent heat 
of vaporization.  Fig.~\ref{fig-bowen} shows
the Bowen ratio as a function of surface temperature, assuming a fixed
surface--air temperature difference and fixed relative humidity. We
have fixed these to values that are representative of the GCM
simulations for simplicity, but the surface--air temperature
difference and relative humidity are not fixed in the GCM.  For
surface temperatures greater than $\about 290 \,\mathrm{K}$, latent
heat fluxes are a factor of $\gtrsim 4$ larger than sensible heat
fluxes, as at Earth's surface \citep{Trenberth09a}.  Similar arguments
apply for net longwave radiative fluxes at the surface, which become
small as the longwave optical thickness of the atmosphere and with it
surface temperatures increase; see \citet{pierrehumbert09} for a more
complete discussion of the surface energy budget.

The precipitation rates also mimic the insolation distribution
(Fig.~\ref{fig-hydro}). 
There is a convergence zone with large precipitation rates ($\geq
40\,\mathrm{mm/day}$) around the subsolar point. Precipitation rates
exceed evaporation rates within $\about 15^\circ$ of the subsolar
point. Outside that region on the day side, evaporation rates exceed
precipitation rates (Fig.~\ref{fig-hydro}), which would lead to the 
generation of deserts there if the surface water supply were limited.  
The atmospheric circulation that gives rise to the moisture transport 
toward the subsolar point is discussed next.

\subsection{General circulation of the atmosphere} 

\begin{figure*}[!tbh]
\centerline{\includegraphics[width=6.0in,clip=true]{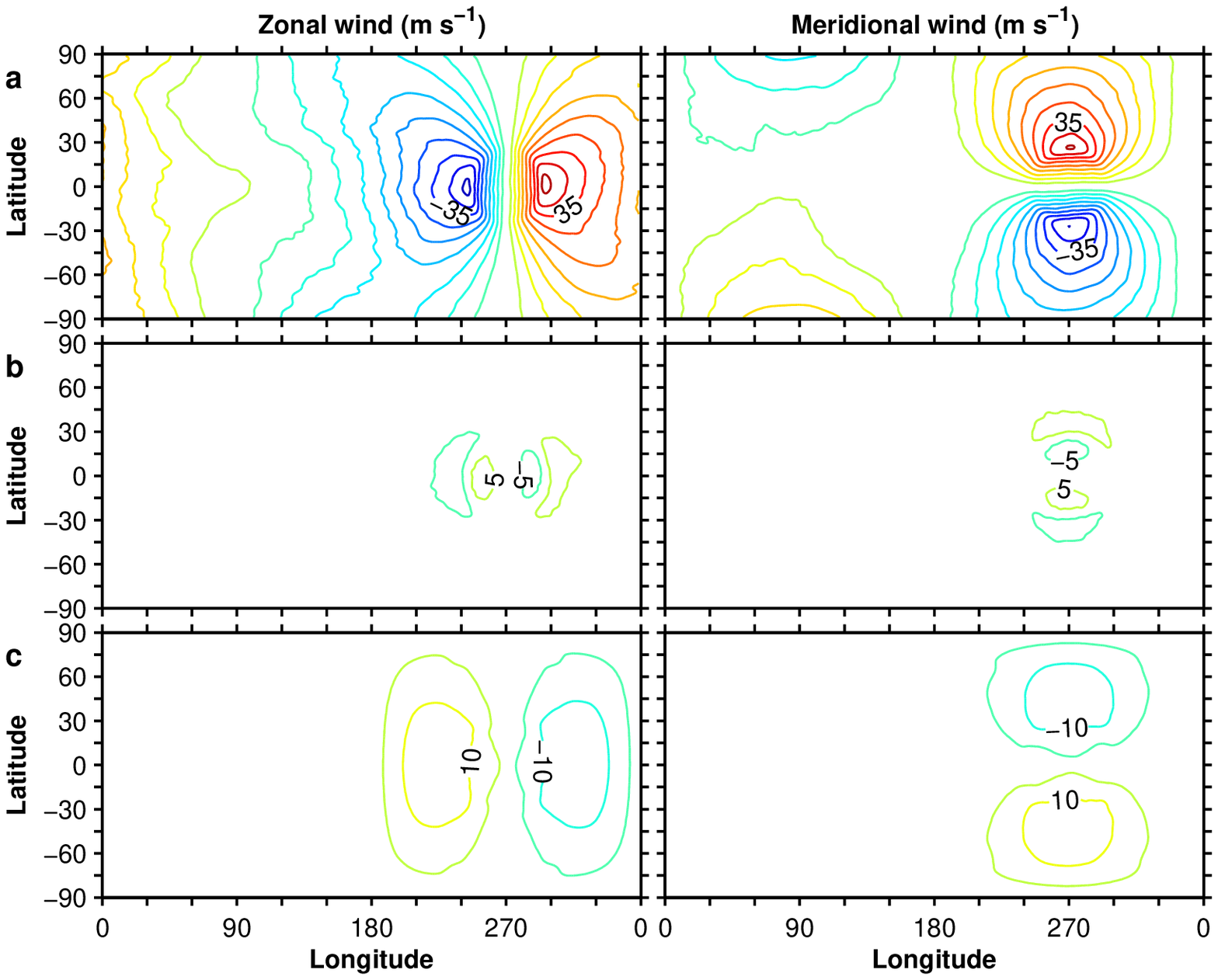}}
\caption{Zonal wind (left) and meridional wind (right) in
  $\Omega_E/365$ simulation on the $\sigma = 0.28$ (a), $0.54$ (b),
  and $1.0$ (c) surfaces. The contour interval is
  $5\,\mathrm{m\,s^{-1}}$.}
\label{fig-winds}
\end{figure*}
\begin{figure*}[!tbh]
\centerline{\includegraphics[width=6.0in,clip=true]{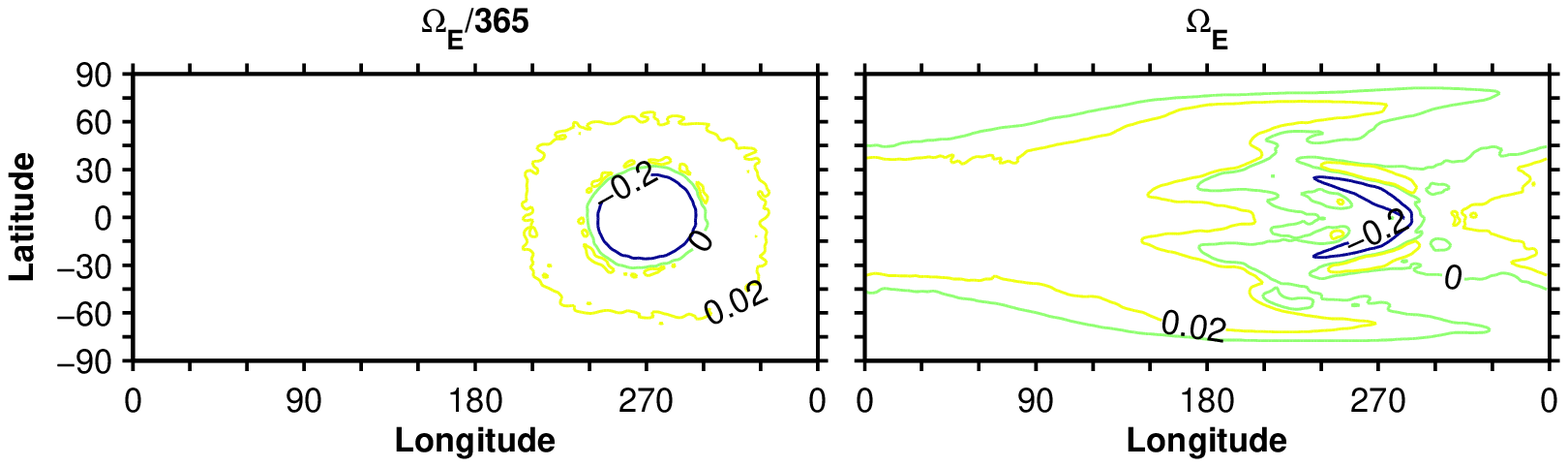}}
\caption{Pressure velocity ($\omega=Dp/Dt$) on the $\sigma = 0.54$
  surface in simulations with $\Omega_E/365$ (left) and $\Omega_E$ (right).
  Contours are shown at $-0.2$, $0$, and
  $0.02\,\mathrm{Pa\,s^{-1}}$.}
\label{fig-omega}
\end{figure*}

The winds are approximately isotropic and divergent at leading
order. They form a thermally direct overturning circulation, with
lower-level convergence near the subsolar point, upper-level
divergence above it, and weaker flow in between
(Fig.~\ref{fig-winds}). As in the simulations of \citet{joshi97}, the
meridional surface flow crosses the poles.  Moist convection in the
vicinity of the subsolar point results in strong mean ascent in the
mid-troposphere there; elsewhere there is weak subsidence associated
with radiative cooling (Fig.~\ref{fig-omega}).  Consistent with the
predominance of divergent and approximately isotropic flow, the Rossby
number is large even in the extratropics ($\Ro \gtrsim 10$).

\begin{figure*}[!tbh]
\centerline{\includegraphics[width=6.0in,clip=true]
{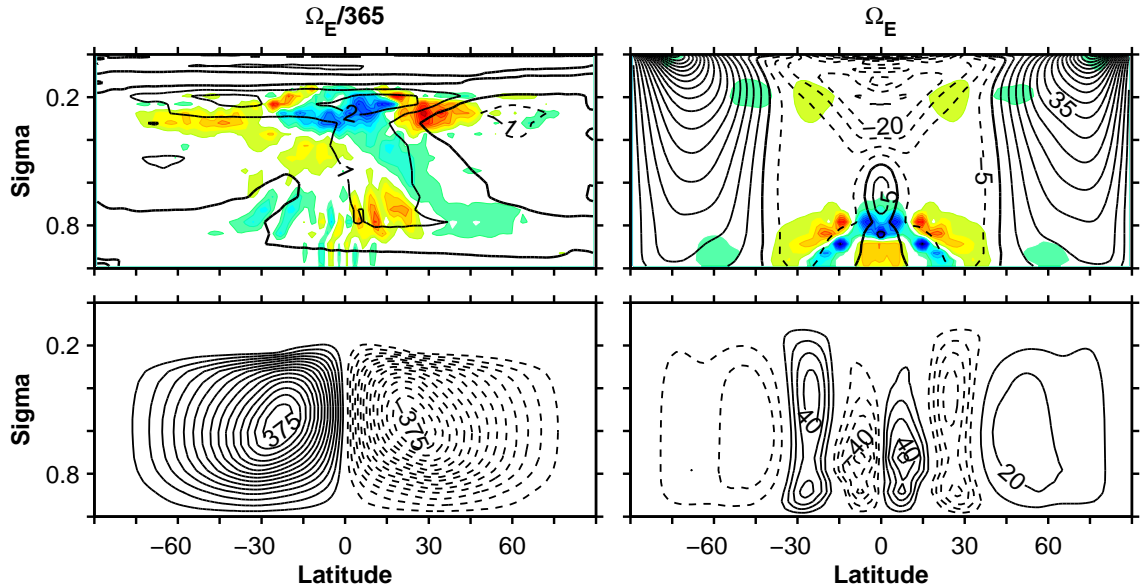}}
\caption{Circulation in simulations with $\Omega_E/365$ (left panels) 
  and $\Omega_E$ (right panels). Top row: zonal mean zonal wind (lines,
  contour interval $1 \, \mathrm{m\,s^{-1}}$ for $\Omega_E$ and
  $5\,\mathrm{m\,s^{-1}}$ for $\Omega_E/365$) and divergence of the
  horizontal component of eddy (transient and stationary) 
  angular momentum fluxes (colors,
  contour interval $1.0 \times 10^{-6}\,\mathrm{m\,s^{-2}}$ for
  $\Omega_E/365$ and $1.5 \times 10^{-5}\,\mathrm{m\,s^{-2}}$ for
  $\Omega_E$). Bottom row: Eulerian-mean mass flux streamfunction
  (contour interval $25 \times 10^9 \,\mathrm{kg\,s^{-1}}$ for
  $\Omega_E/365$ and $10 \times 10^9 \,\mathrm{kg\,s^{-1}}$ for
  $\Omega_E$).}
\label{fig-zonalmean_circ}
\end{figure*}
The zonal mean zonal wind and streamfunction are shown in
Fig.~\ref{fig-zonalmean_circ}. The zonal mean zonal wind is a weak
residual of the opposing contributions from different longitudes
(Fig.~\ref{fig-winds}). In the upper troposphere near the equator,
there are weak westerly (superrotating) zonal winds, as well as the
eddy angular momentum flux convergence that, according to Hide's
theorem, is necessary to sustain them
\citep[e.g.,][]{hide69,schneiderEK77b,held80,schneider06b}.

The Eulerian mean mass streamfunction consists of a Hadley cell in
each hemisphere, which is a factor $\about 3$ stronger than Earth's
and extends essentially to the pole. These Hadley cells are
thermodynamically direct circulations: the poleward flow to higher
latitudes has larger moist static energy ($h = c_p T + g z + L q$) 
than the near-surface return
flow. The classic theory of Hadley cells with nearly inviscid, angular
momentum-conserving upper branches predicts that the Hadley cell
extent increases as the rotation rate decreases
\citep{schneiderEK77b,held80}. However, the results of this theory do
not strictly apply here because several assumptions on which it is
based are violated. For example, the surface wind is not weak relative
to the upper-tropospheric winds (Fig.~\ref{fig-winds}), and the zonal
wind is not in balance with meridional geopotential (or temperature)
gradients (i.e., the meridional wind is not negligible in the
meridional momentum equation).
Nonetheless, it is to be expected that Hadley cells in
slowly rotating atmospheres span hemispheres \citep{williams88a}.

\begin{figure*}[!tbh]
  \centerline{
    \includegraphics[width=5.4in,clip=true]{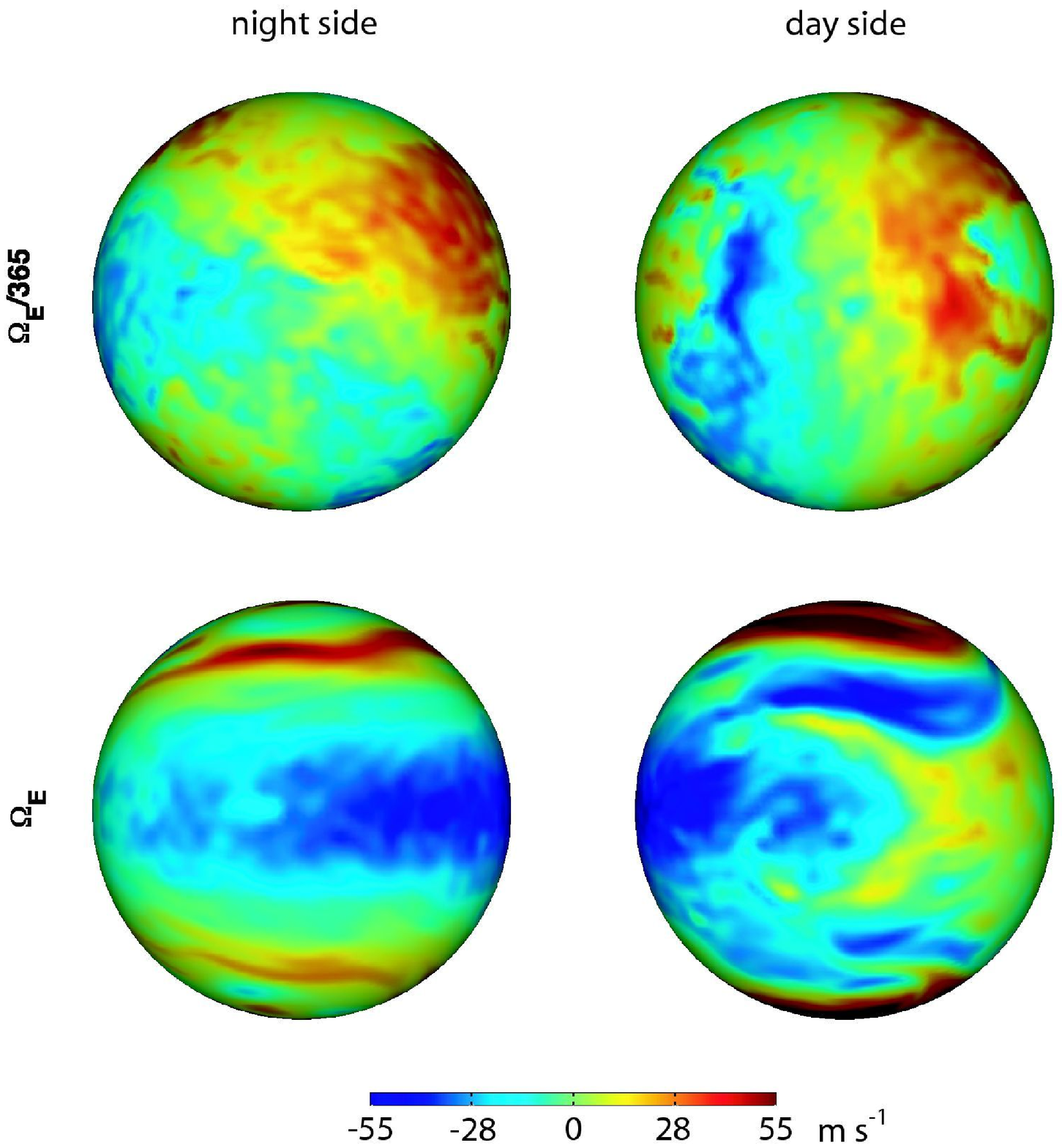}
  }
  \caption{Instantaneous zonal wind on $\sigma = 0.28$ surface in the
    statistically steady state (on day 3900) of the two
    simulations. Animations are available at \slow_anim and \fast_anim .}
  \label{fig-instantaneous_zonalwind}
\end{figure*}
The instantaneous, upper-tropospheric zonal wind is shown in
Fig.~\ref{fig-instantaneous_zonalwind}, and a corresponding animation
is available at \slow_anim.
Athough the flow statistics in the simulations
are hemispherically symmetric (because the forcing and boundary
conditions are), the instantaneous wind exhibits north-south
asymmetries on large scales and ubiquitous variability on smaller
scales. The large-scale variability has long timescales
($\about 80$~days), and the zonal-mean zonal wind is
sensitive to the length of the time averaging: for timescales as long as
the 500~days over which we averaged, the averages still exhibit
hemispheric asymmetries (Fig.~\ref{fig-zonalmean_circ}).

\begin{figure*}[!tbh]
\centerline{\includegraphics[width=6.0in,clip=true]
{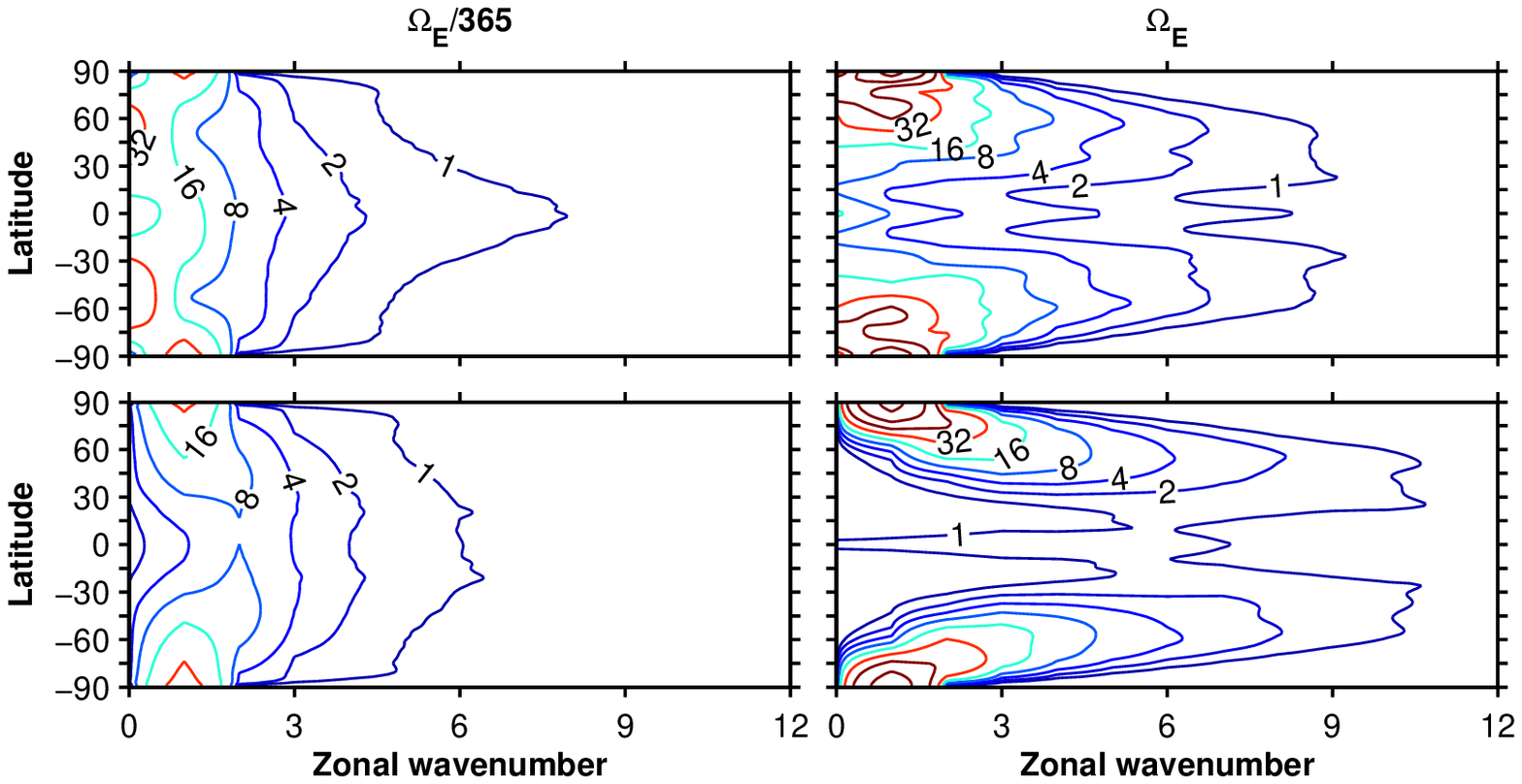}}
\caption{Spectrum of the mass-weighted, vertically averaged transient
  velocity variance for zonal (top row) and meridional (bottom row)
  wind components in the simulations with $\Omega_E/365$ (left) and 
  $\Omega_E$ (right). Contours are shown at $2^0, 2^1, ..., 2^5 \,
  \mathrm{m^2\,s^{-2}}$.}
\label{fig-vel_var_spec}
\end{figure*}

The vertically integrated eddy kinetic energy is $8.4 
\times 10^5 \ \mathrm{J \ m^{-2}}$ in the global mean.  
The instantaneous velocity fields
exhibit substantial variability on relatively small scales
(e.g., Fig.~\ref{fig-instantaneous_zonalwind}), and the kinetic energy
spectrum decays only weakly from spherical wavenumber $\about 20$
toward the roll-off near the grid scale owing to the subgrid-scale
filter. However, the peak in velocity variance is at the largest
scales (Fig.~\ref{fig-vel_var_spec})---as suggested by the
hemispherically asymmetric large-scale variability seen in
Fig.~\ref{fig-instantaneous_zonalwind} and in the animation.

\subsection{Atmospheric stratification and energy transports}

\begin{figure*}[!tbh]
\centerline{\includegraphics[width=6.0in,clip=true]
{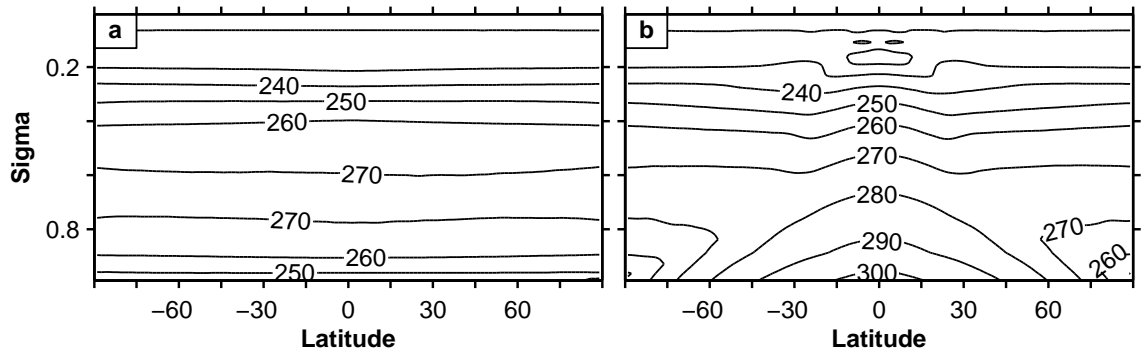}}
\caption{Temperature section of antisolar longitudes (a) and subsolar
  longitudes (b) of $\Omega_E/365$ simulation.  Averages are taken
  over 10$^\circ$ of longitude.  The contour interval is
  $10\,\mathrm{K}$.}
\label{fig-temp}
\end{figure*}

\begin{figure*}[!tbh]
\centerline{\includegraphics[width=6.0in,clip=true]
{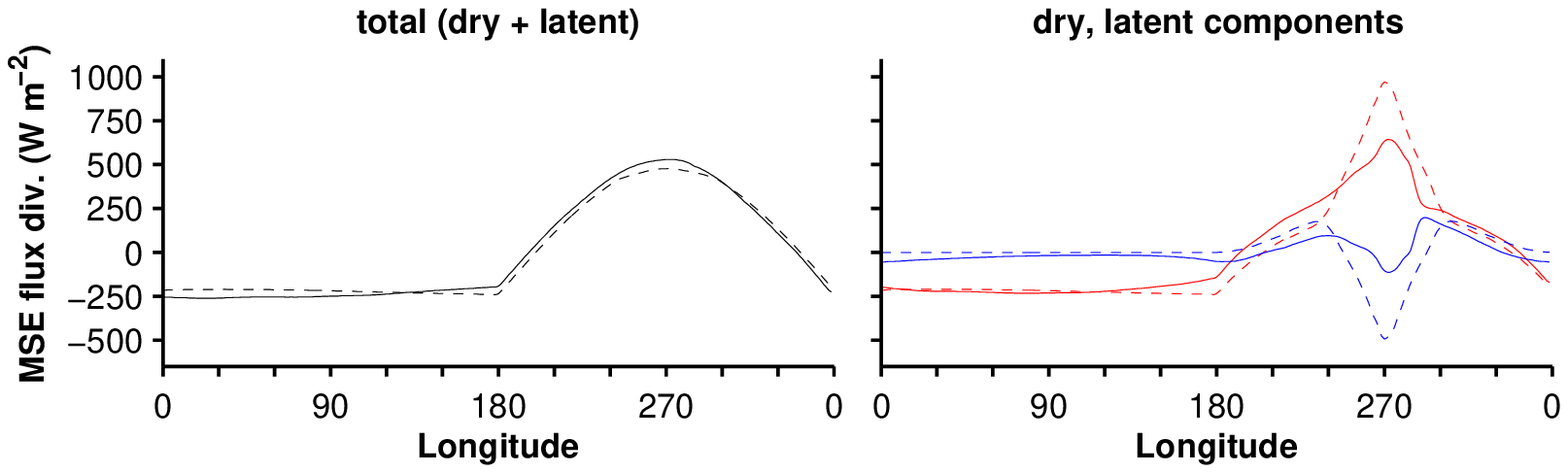}}
\caption{Left: Vertical and meridional integral of the divergence of 
  the moist static energy flux, $u h$. Right: Dry static energy (red) and 
  latent energy (blue) components of moist static enery flux divergence. 
  Dashed lines for $\Omega_E/365$ simulation and solid lines for $\Omega_E$
  simulation.}
\label{fig-mse_fluxes}
\end{figure*}
Consistent with the scaling arguments presented above, horizontal
variations in temperature are small in the free troposphere (above the
$\about 750$~hPa level, Fig.~\ref{fig-temp}). This
is reminiscent of Earth's tropics, where temperature variations are
small because Coriolis accelerations are weak compared with inertial
accelerations \citep[e.g.,][]{charney63,sobel01}.  Here, Coriolis
accelerations are weak at all latitudes, and free-tropospheric
horizontal temperature variations are small everywhere.

The thermal stratification on the day side in the simulation is moist
adiabatic in the free troposphere within $\about 30^{\circ}$ of the
subsolar point. Farther away from the intense moist convection around 
the subsolar point, including on the night side of the planet, there are
low-level temperature inversions (Fig.~\ref{fig-temp}b), as in the
simulations in \citet{joshi97}. These regions have small optical thickness
because of the low water vapor concentrations and therefore have strong
radiative cooling to space from near the surface.
The inversions arise because horizontal temperature gradients are
constrained to be weak in the free troposphere, but near-surface air
cools strongly radiatively, giving rise to inversions.

The vertical and meridional integral of the horizontal moist static
energy flux divergence as a function of longitude is shown in 
Fig.~\ref{fig-mse_fluxes} (dashed curves). Energy diverges on the day 
side of the planet 
and converges on the night side of the planet, reducing temperature 
contrasts. Near the subsolar point, there is substantial cancellation
between the latent energy and dry static energy components of the moist static
energy flux divergence as there is, for example, in Earth's Hadley 
circulation. As can be inferred from the 
minimal water vapor flux divergence on the night side
(Fig. ~\ref{fig-hydro}), the moist static energy flux divergence is
dominated by the dry static energy ($c_p T + g z$) component on the
night side ($\gtrsim 99 \%$ of the total). 

\section{Rapidly rotating simulation} \label{section-rapid}

In rapidly rotating atmospheres, the Rossby number in the extratropics
is small, and geostrophic balance is the dominant balance in the 
horizontal momentum equations.
In the zonal mean, zonal pressure gradients vanish, but meridional pressure
gradients do not, so if the dominant momentum balance is geostrophic,
winds are anisotropic $\overline{U} > \overline{V}$, 
where $\overline{(\cdot)}$ denotes a zonal mean. 
Variations in the planetary vorticity with latitude, $\beta$, are 
central to vorticity mixing arguments \citep[e.g.,][]{rhines94,held00b}, 
which can account for the generation of atmospheric jets: When a Rossby 
wave packet stirs the atmosphere in 
a region bounded by a polar cap, high-vorticity fluid moves equatorward and 
low-vorticity fluid moves poleward. This reduces the vorticity in the
polar cap. By Stokes' Theorem, the reduced vorticity in the polar cap 
means the zonal wind at the latitude of the bounding cap decreases;
if angular momentum is conserved, the zonal wind outside the polar 
cap increases. Irreversible vorticity mixing (wave-breaking or
dissipation) is necessary to maintain the angular momentum fluxes in the
time mean. Thus, the larger planetary vorticity gradients
of the rapidly rotating planet allow jets to form provided there is a source
of wave activity that leads to vorticity stirring.

We return to the analysis of \citet{charney63} to estimate temperature
variations in the free atmosphere. For rapidly rotating planets, 
geostrophic balance holds in the horizontal momentum equations: 
$\delta p / (\rho L) \sim f U$.
Combining the scaling from the momentum equation with the hydrostatic 
relation,
the pressure, density, and (potential) temperature variations scale like the 
ratio of the Froude number to the Rossby number,
\begin{equation}
\frac{\delta p }{p} \sim \frac{\delta \rho }{\rho} \sim 
\frac{\delta \theta }{\theta} \sim \frac{f U L}{g H} = \frac{\Fr}{\Ro}.
\end{equation}
Where the Rossby number is small (in the extratropics), 
the temperature variations will be a factor of 
order inverse Rossby number ($\Ro^{-1} \sim 10$) larger than in the slowly 
rotating simulation if similar values for $U, g,$ and $H$ are assumed. 
Thus, we expect larger horizontal temperature and pressure variations 
away from the equator in the rapidly rotating simulation.

\subsection{Surface temperature}

In the rapidly rotating simulation, the surface temperature
on the day side of the planet is maximal off of the
equator and does not bear a close resemblance to the insolation
distribution; the night side of the planet has relatively warm
regions in western high latitudes (Fig.~\ref{fig-surf_temp}).
Compared to the slowly rotating simulation, the surface temperature is
more substantially modified by the atmospheric circulation; however,
the temperature contrasts between the day and night side are similar.

\subsection{Hydrological cycle}

Surface evaporation rates mimic the insolation distribution 
(Fig.~\ref{fig-hydro}). This is one of the most similar fields between
the slowly and rapidly rotating simulations, as expected from the gross 
similarity in surface temperature and the smallness of the Bowen ratio at
these temperatures (Fig.~\ref{fig-bowen}). This might not be the case if 
the model included the radiative effects of clouds since the amount of 
shortwave radiation reaching the surface would be shaped by variations in 
cloud albedo which, in turn, depend on the atmospheric circulation.

Precipitation rates are large in a crescent-shaped region
on the day side of the planet; the night side of the
planet generally has small but nonzero precipitation rates
(Fig.~\ref{fig-hydro}).  The evaporation minus precipitation field
has substantial structure: there are large amplitude changes from the
convergence zones ($P > E$) to nearby
areas of significant net drying ($E > P$).  Comparing the slowly and
rapidly rotating simulations shows that precipitation and $E -
P$ on the night side of the planet are sensitive to the atmospheric
circulation.

An interesting aspect of the climate is that the maximum precipitation
(near $\about 15^{\circ}$ latitude) is not co-located with the 
off-equator maximum surface temperature (near $\about 40^{\circ}$ latitude). 
The simulation provides an example of precipitation and 
deep convection that are not locally thermodynamically controlled: 
the precipitation is not maximum where the surface temperature is maximum;
the column static stability (e.g., Fig.~\ref{fig-temp_earthomega}), 
and therefore the convective available potential energy, are not markedly 
different between the 
maxima in precipitation and temperature. However, if the surface climate is 
examined latitude-by-latitude instead of examining its global maxima, the 
region of large precipitation is close to the maximum surface temperature (as
well as surface temperature curvature) at a given latitude. 
The structure of the surface winds, discussed next, and the associated 
moisture convergence are key for determining where precipitation is large.
\citet{sobel07a} provides a review of these two classes of theories 
for tropical precipitation (thermodynamic control vs. momentum control) 
and the somewhat inconclusive evidence
of which class of theory better accounts for Earth observations.

The global precipitation is $\sim$10\% larger in the rapidly rotating 
simulation than in the slowly rotating simulation. This suggests that 
radiative-convective equilibrium cannot completely describe the strength
of the hydrological cycle.

\subsection{General circulation of the atmosphere}

\begin{figure*}[!tbh]
\centerline{\includegraphics[width=6.0in,clip=true]{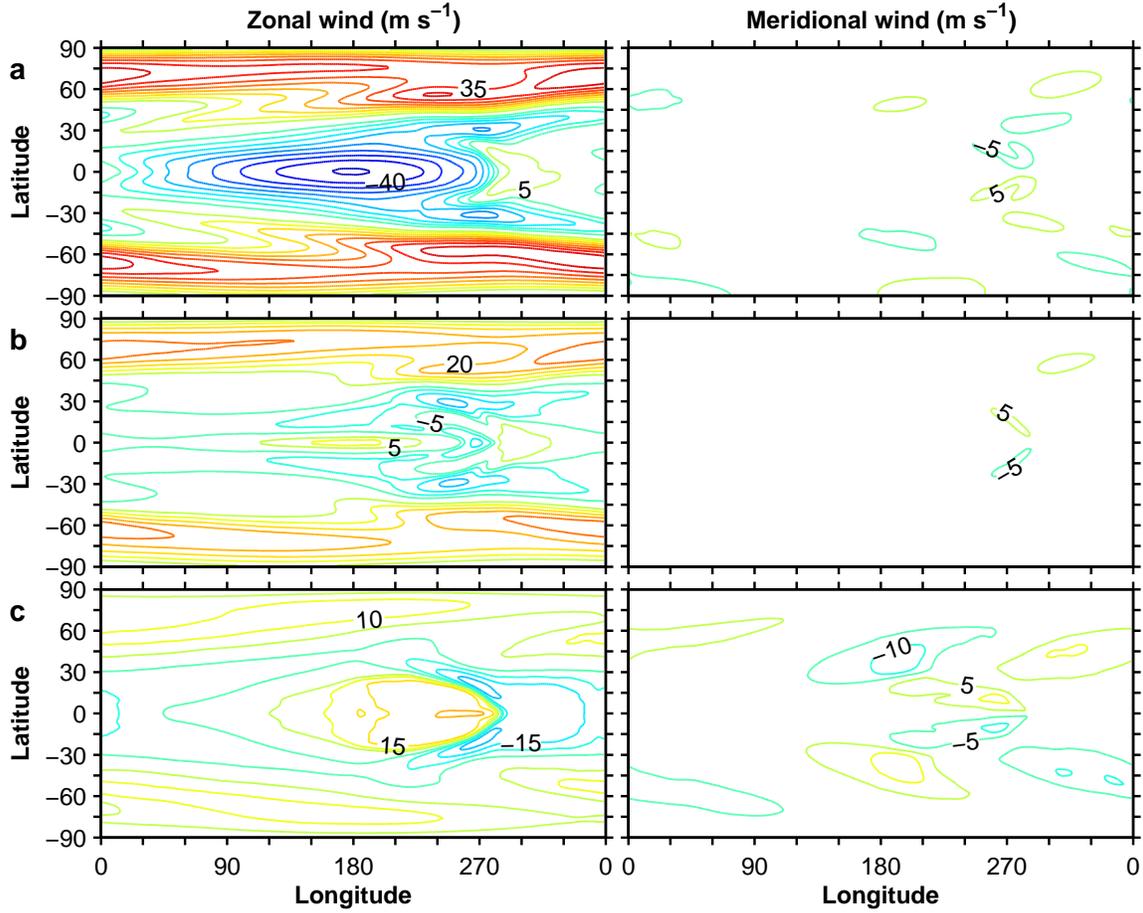}}
\caption{Zonal wind (left) and meridional wind (right) of $\Omega_E$
  simulation on the $\sigma = 0.28$ (a), $0.54$ (b), and $1.0$ (c)
  surfaces. The contour interval is   $5\,\mathrm{m\,s^{-1}}$. }
\label{fig-winds_earthomega}
\end{figure*}

In the rapidly rotating simulation, the atmospheric circulation has
several prominent features: there are westerly jets in high 
($\about 65^{\circ}$) latitudes, the mid-tropospheric zonal wind
exhibits superrotation, and the surface winds converge in a crescent-shaped
region near the subsolar point (Fig.~\ref{fig-winds_earthomega}). 
The equatorial superrotation and westerly jets are clear in the
zonal mean zonal wind (Fig.~\ref{fig-zonalmean_circ}).

\begin{figure*}[!tbh]
\centerline{\includegraphics[width=6.0in,clip=true]
{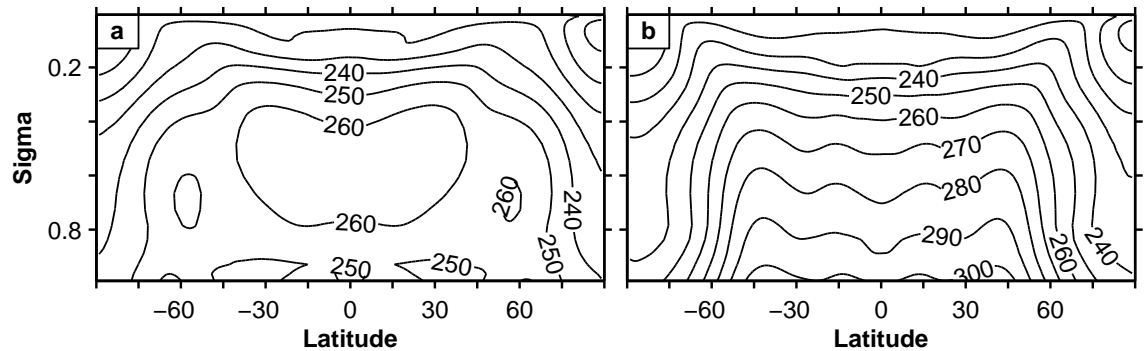}}
\caption{Temperature section of antisolar longitudes (a) and subsolar 
  longitudes (b) of $\Omega_E$ simulation. 
  Averages are taken over 10$^\circ$ of longitude.
  The contour interval is $10\,\mathrm{K}$.}
\label{fig-temp_earthomega}
\end{figure*}

The existence of the high-latitude jets can be understood from the 
temperature field and eddy angular momentum flux convergence 
(Figs.~\ref{fig-temp_earthomega} and ~\ref{fig-zonalmean_circ}).
There are large meridional temperature gradients, which give rise to 
zonal wind shear by thermal wind balance and provide available potential
energy for baroclinic eddies that transport angular momentum into the 
jets. In the vertical average, the eddy angular momentum transport into
an atmospheric column is balanced by surface stress.
Note that the region of higher temperatures on the night side of the
planet coincides with the westerly jet's surface maximum.
The stronger winds at these latitudes lead to strong temperature advection.
The equatorial superrotation is a consequence of angular momentum flux
convergence (Fig.~\ref{fig-zonalmean_circ}). \citet{saravanan93} and 
\citet{suarez92} describe the emergence of superrotation 
generated by large-scale, zonally-asymmetric heating anomalies in the tropics.
As in their idealized models, the zonal asymmetry in the low-latitude heating
(in our simulation, provided by insolation) generates a stationary Rossby wave.
Consistent with a stationary wave source,
in the rapidly rotating simulation, the horizontal eddy angular momentum flux 
convergence in low latitudes is dominated by the stationary eddy component.
This aspect of the simulation is sensitive to horizontal resolution and
the subgrid-scale filter. With higher resolution or weaker filtering, the 
superrotation generally extends higher into the troposphere and has a 
larger maximum value.

\begin{figure*}[!tbh]
\centerline{\includegraphics[width=6.0in,clip=true]
{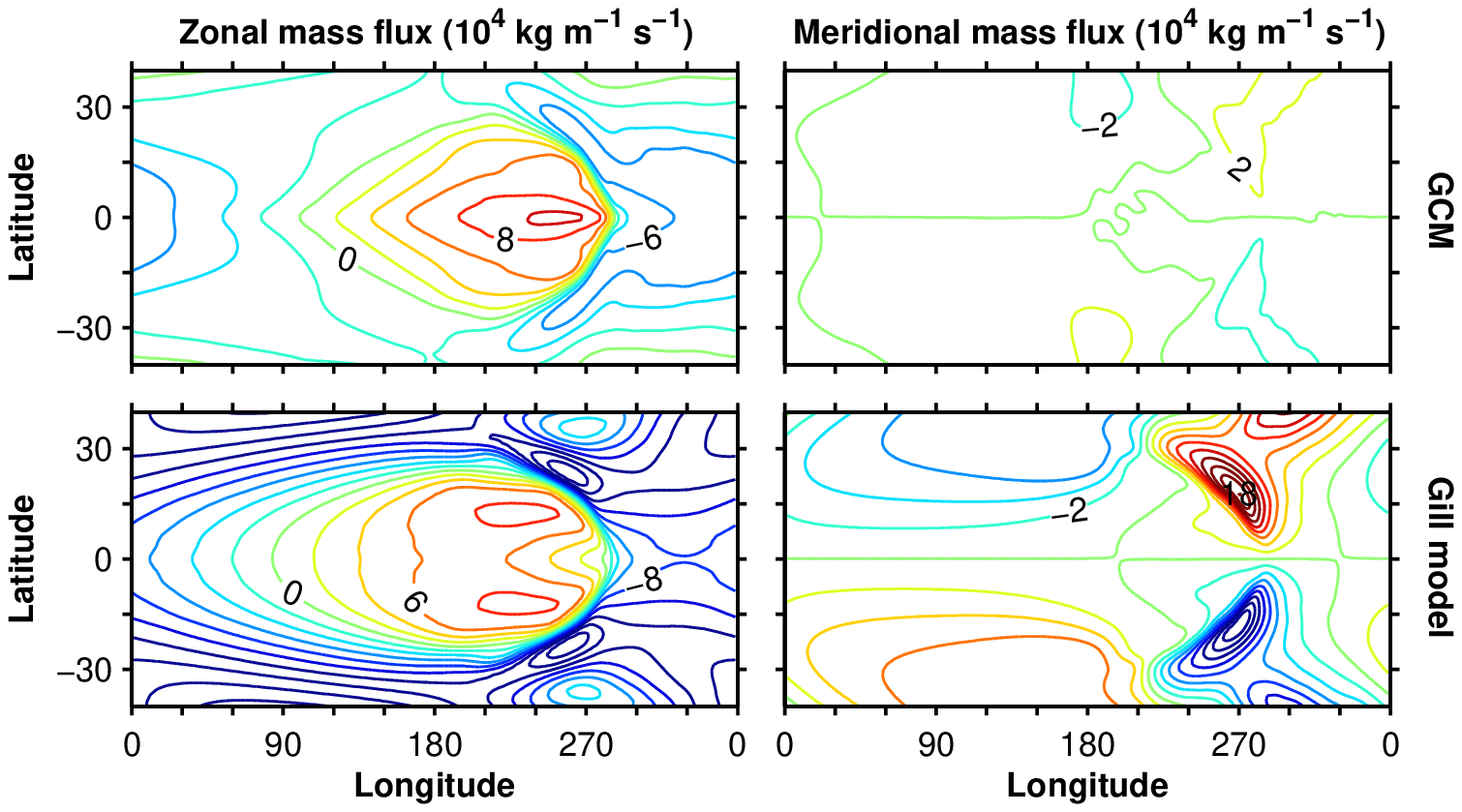}}
\caption{Zonal (left column) and meridional (right column) 
  near-surface mass fluxes for the GCM 
  (top row, averaged between the surface and $\sigma = 0.73$ model level)
  and the Gill model forced by the GCM's precipitation (bottom row,
  see Appendix for Gill model equations and parameters).
  The contour interval is  $2 \times 10^4\,\mathrm{kg\,m^{-1}\,s^{-1}}$.}
\label{fig-sfc_massfluxes}
\end{figure*}

There is a crescent-shaped region where the surface zonal wind is converging.
This is where the precipitation (Fig.~\ref{fig-hydro}) and upward vertical 
velocity (Fig.~\ref{fig-omega}) are largest.
The horizontal scale of the convergence zone is similar to the equatorial 
Rossby radius, $(c/\beta)^{1/2} \sim 10^{\circ}$, where $\beta$ is the 
gradient of planetary vorticity and $c$ is the gravity wave speed (estimated
using a characteristic tropospheric value for the Brunt-V\"{a}is\"{a}l\"{a}
frequency on the day side of the planet).
The surface zonal wind can be qualitatively understood as the 
equatorially-trapped wave response to stationary heating: 
equatorial Kelvin waves propagate to the east of the heating and 
generate easterlies;
equatorial Rossby waves propagate to the west of the heating and generate
westerlies \citep{gill80}.

The shape of the zero zonal wind line and its horizontal scale 
are similar to those of the \citet{gill80} model, which 
describes the response of damped, linear shallow-water waves to
a prescribed heat/mass source.
For the prescribed heat source in the original Gill model, the crescent-shape 
zero zonal wind line extends over $\sim$2 Rossby radii and, as in the 
GCM simulation, is displaced to the east of maximum heating on the equator. 

A complicating factor in the analogy between the GCM's low-latitude
surface winds and those of the Gill model is that the heating is prescribed
in the Gill model, while it is interactive with the flow in the GCM.
As previously mentioned, the precipitation is strongly shaped by the winds.
To see if the analogy between the winds in the GCM and in the
Gill model breaks down 
because of the more complex structure of the latent heating,
we force a variant of the Gill model with the GCM's precipitation field
following the formulation of \citet{neelin88} (see Appendix for details). 
The results of this calculation
are compared with the GCM output in Fig.~\ref{fig-sfc_massfluxes}.
The direction, large-scale structure, and, in the case of the zonal 
component, magnitude of the mass fluxes are similar between the GCM and 
precipitation-forced Gill model, though it is clear that there are 
quantitative differences.

In contrast to the larger-magnitude zonal wind,
the meridional wind is diverging at the surface and converging
aloft near the subsolar point
(right panel of Fig.~\ref{fig-winds_earthomega}a,c). 
The dynamics of this are contained in the inviscid limit of the linear 
shallow-water equations: for a Sverdrup vorticity
balance, the vortex stretching caused by the overall convergence 
near the surface near the equator must be balanced by moving poleward,
toward higher planetary vorticity; hence the poleward meridional wind
near the surface and the equatorward meridional wind aloft.

The Eulerian mean mass streamfunction (Fig.~\ref{fig-zonalmean_circ}) has the
opposite sense as Earth's Hadley cells:
in the zonal mean, there is descent on the equator, poleward flow 
near the surface, ascent near $15^{\circ}$, and equatorward flow in the 
mid-troposphere.
If the dominant balance in the zonal momentum equation is between 
Coriolis acceleration and eddy angular momentum flux divergence, 
$-f \overline{v} \sim - \partial_y \overline{u'v'}$ (i.e., small 
local Rossby number as defined in \citet{walker06}), then
the angular momentum flux convergence that establishes the superrotating 
zonal wind
also leads to equatorward mean meridional wind in the free troposphere,
as in Earth's Ferrel cells \citep[e.g.,][]{held00b}. 
This can lead to a dynamical feedback that enhances superrotation 
\citep{schneider09a}: as superrotation emerges, the mean meridional 
circulation can change direction with a concomitant change in the 
direction of mean-flow angular momentum fluxes (changing from exporting 
angular momentum from the deep 
tropics to importing it), which enhances the superrotation.

The instantaneous, upper tropospheric zonal wind in the rapidly rotating
simulation is shown in Fig.~\ref{fig-instantaneous_zonalwind}, and a 
corresponding animation is available at \fast_anim.
The large-scale features of the general circulation such as the high-latitude 
jets and divergent zonal wind in the tropical upper troposphere are clear in 
the instantaneous winds. The eddies in the animation generally have larger 
spatial scales and longer timescales than in the corresponding slowly rotating
animation. 

The vertically and globally integrated eddy kinetic energy 
is $1.0 \times 10^6 \ \mathrm{J \ m^{-2}}$. This is about 20\% larger than 
int the slowly rotating case. 
The eddy kinetic energy spectrum has a typical, steep $n^{-3}$ 
shape to the smallest resolved wavenumber. 
In some sense, the integrated eddy kinetic energy
belies the difference in synoptic variability between the rapidly and slowly 
rotating simulations: in the extratropics, for zonal wavenumbers between 
$\sim$3-6, the rapidly rotating simulation has a factor of 2-3 times more
velocity variance (Fig.~\ref{fig-vel_var_spec}) than the slowly rotating
simulation.

\subsection{Atmospheric stratification and energy transports}

The tropospheric temperatures on the day side of the planet 
(Fig.~\ref{fig-temp_earthomega}) reflect the surface temperature 
variations: the temperature field has a local
maximum near $\sim$40$^{\circ}$ and is relatively uniform up to high 
latitudes (poleward of $\sim$50$^{\circ}$). In the free troposphere on the
day side, the lapse rates are close to the moist adiabatic lapse rate, 
computed using the local temperature and pressure, over a region 
roughly within the $300\, \mathrm{K}$ contour of the surface 
temperature. Note that $300\, \mathrm{K}$ does not have a particular 
physical significance---we are simply using it to describe a feature
of the simulation. 
There is a local minimum of temperature on the equator which 
may be related to the downward vertical velocity there (Fig.~\ref{fig-omega}).
In low latitudes, there is a near-surface inversion on the night side of 
the rapidly rotating 
simulation that, as in the slowly rotating simulation, is the result of
weak temperature gradients in the free troposphere and the small optical
thickness of the atmosphere.

As in the slowly rotating simulation, the moist static energy flux divergence
is positive on the day side and negative on night side of the planet 
(solid curves in Fig.~\ref{fig-mse_fluxes}); there is substantial
cancellation between dry static energy and latent energy flux divergence
near the subsolar point.
Though the hydrological cycle is more active on the night side of the 
rapidly rotating simulation than in the slowly rotating simulation, 
the dry static energy component still dominates ($\gtrsim 80 \%$) the moist 
static energy advection on the night side of the planet.

The two simulations are more similar in this respect than might have been
anticipated given the differences in their flow characteristics, although
there are regional differences that are obscured by averaging over latitude.
The broad similarities can be understood by considering the moist static
energy budget. In the time mean, denoted $\overline{(\cdot)}$, 
neglecting kinetic energy divergence and diffusive
processes within the atmosphere, the mass-weighted vertical integral, 
denoted $\langle \cdot \rangle$, of the moist static energy flux divergence
is balanced by surface energy fluxes, $F_\mathrm{s}$, 
and radiative tendencies, $Q_{\mathrm{rad}}$: 
\begin{equation}
\nabla \cdot \langle \overline{ {\bf u} h} \rangle = 
\overline{F}_\mathrm{s} + \overline{Q}_{\mathrm{rad}}.
\end{equation}
As a result of the gross similarity of the two simulations in evaporation and 
low-latitude stratification (in part due to the smaller dynamical role that
rotation plays near the equator)
and hence the radiative cooling, the divergence of the moist static energy 
flux is also similar. 
In the extratropics, considering the sources and sinks of moist static
energy does not provide a useful constraint because the
stratification is dynamically determined \citep[e.g.,][]{held78c, schneider07b}
and there can be geostrophically balanced temperature gradients in rapidly
rotating atmospheres; as result, the radiative cooling, through these 
dynamical influences on temperature, is determined by the flow,
so the moist static energy flux divergence will not in general be 
independent of rotation rate. Indeed, the warm regions on the night side of 
the rapidly rotating simulation have larger moist static energy flux divergence
than in those regions on the night side of the slowly rotating simulation 
as a result of the difference in circulation.

\section{Conclusions}

We have examined the dynamics of tidally locked, Earth-like planets.
In particular, the dynamical regime of the atmosphere depends on the planet 
rotation rate, as anticipated from scaling arguments.
The simulations demonstrate the importance of the atmospheric circulation in 
determining the surface climate. 
For example, the difference between the precipitation distributions in 
the slowly and rapidly rotating simulations clearly shows the dependence 
of the hydrological cycle on the circulation of the atmosphere.
Interestingly, some aspects of the simulated climate are not sensitive to
the planet's rotation rate. In particular, the temperature contrast between
the day and night side and evaporation rates are similar between the slowly
and rapidly rotating simulations.

The general circulation of the slowly rotating atmosphere
features global-scale, thermodynamically-direct, divergenct circulations 
with ascending motion at the subsolar point and descending motion elsewhere.
In contrast, the general circulation of the rapidly rotating atmosphere
features extratropical jets that owe their existence to rotational 
(Rossby) waves, and tropical surface winds that are the result of 
stationary equatorial (Rossby and Kelvin) waves.
The isotropy of the winds in the slowly rotating regime and the anisotropy
of the winds in the rapidly rotating regime are expected from the dominant
balance of the horizontal momentum equations.
Expectations for the free-atmospheric pressure and temperature gradients
based on scale analysis were realized in the simulations of the different 
circulations regimes: temperature gradients are weak where Coriolis 
accelerations are weak (in low latitudes of the rapidly rotating atmosphere 
and globally in the slowly rotating atmosphere) and larger where 
Coriolis accelerations are dynamically important (in the extratropics 
of the rapidly rotating atmosphere).

While many aspects of the simulations can be explained by 
general circulation theories for Earth and Earth-like atmospheres, there 
remain quantitative questions that require more systematic experimentation 
than we have attempted here. 
For example, the dependence of the surface climate on the solar constant
may be different from simulations with Earth-like insolation because 
of the differences in stratification.

\bigskip
{\it Acknowledgments}. Timothy Merlis was supported by a National Defense
Science and Engineering Graduate fellowship and a National Science Foundation
Graduate Research fellowship.
We thank Sonja Graves for providing modifications to the GCM code and
Simona Bordoni, Ian Eisenman, Andy Ingersoll, and Yohai Kaspi
for comments on a draft of the manuscript.
The simulations were performed on Caltech's Division of Geological and 
Planetary Sciences Dell cluster. The program code for the simulations, 
based on the Flexible Modeling System of the Geophysical Fluid Dynamics 
Laboratory, and the simulation results themselves are available from the 
authors upon request. 

\section{Appendix: Gill model forced by precipitation}

We use the formulation for a Gill model forced by 
precipitation \citep{neelin88}, as opposed to the standard mass sink.
The model equations are
\begin{align}
-f \tilde{v} &= - \partial_x \tilde{\phi} - \epsilon_m \tilde{u} \\
f \tilde{u} &=  - \partial_y \tilde{\phi} - \epsilon_m \tilde{v} \\
c^2 ( \partial_x \tilde{u} + \partial_y \tilde{v} ) &= - a P 
- \epsilon_T \tilde{\phi},
\end{align}
where $a = L R / c_p$ is the combination of the latent heat of vaporization,
gas constant, and heat capacity at constant pressure that results from 
converting the precipitation rate to a temperature tendency in the 
thermodynamic equation and approximating Newtonian cooling by geopotential
damping,
$\epsilon_m$ is the Rayleigh drag coefficient of the momentum equations,
$\epsilon_T$ is the damping coefficient for the geopotential,
other variables have their standard meanings (e.g. $P$ is precipitation),
and $\tilde{ (\cdot) } = \int_{p_b}^{p_s} \ (\cdot) \ dp/g$ is the 
mass-weighted vertical integral over the boundary layer ($p_b$ is the pressure
at the top of the boundary layer). We use fairly standard values for
the constants (Table ~\ref{table-constants}). The qualitative aspects 
of the solutions that we discuss are not sensitive to these choices.

\begin{table*}
\begin{tabular}{c c}
\hline \hline
\rule{0pt}{2.5ex} constant & value \\ [.5ex]
\hline
c & 50 m s$^{-1}$ \\
$\beta$ & 2.22 $\times 10^{-11}$ m$^{-1}$ s$^{-1}$ \\
$\epsilon_m$ & (1 day)$^{-1}$ \\
$\epsilon_T$ & (20 day)$^{-1}$ \\
a & 7.14 $\times 10^5$ J kg$^{-1}$ \\
\hline
\end{tabular}
\caption{Gill model constants and their values.}
\label{table-constants}
\end{table*}

The three equations can be combined into a single equation for
$\tilde{v}$:
\begin{equation}
\begin{split}
[-\epsilon_T (\epsilon_m^2 + \beta^2 y^2 )/c^2  + 
\epsilon_m (\partial_{xx} + \partial_{yy}) + \beta \ \partial_x] \tilde{v} = \\
- (\epsilon_m \partial_y - \beta y \ \partial_x) \ a P/c^2,
\end{split}
\end{equation}
where $\beta$-plane geometry ($f = \beta y$) has been assumed.
The boundary value problem for $\tilde{v}$ is
discretized by Fourier tranforming in the x-direction and
finite differencing in the y-direction. It is solved using the GCM's 
climatalogical precipitation field on the right-hand side
(Fig.~\ref{fig-hydro}).

\bigskip
\medskip
{\small
\bibliographystyle{ametsoc}
\bibliography{tidally_locked}

}

\end{multicols}

\end{document}